\documentclass[aps,eqsecnum,nofootinbib,showpacs]{revtex4}
\usepackage{amsmath,amsfonts,amssymb,bm}

\newcommand{\be}{\begin{equation}}
\newcommand{\ee}{\end{equation}}
\newcommand{\bea}{\begin{eqnarray}}
\newcommand{\eea}{\end{eqnarray}}

\newcommand{\eq}[1]{Eq.~(\ref{eq:#1})}
\newcommand{\sect}[1]{Sec.~\ref{sec:#1}}
\newcommand{\appen}[1]{App.~\ref{sec:#1}}

\newcommand{\del}{\partial}
\newcommand{\eT}{\epsilon_T}
\newcommand{\tilG}{\tilde{G}}

\newcommand{\vecx}{\vec{x}}

\newcommand{\vecq}{\vec{q}}
\newcommand{\bareta}{\bar{\eta}}

\numberwithin{equation}{section}
\newcommand{\calA}{\mathcal{A}}
\newcommand{\calL}{\mathcal{L}}

\newcommand{\calV}{\mathcal{V}}
\bmdefine{\bmg}{{\bm{g}}}
\bmdefine{\bmA}{{\bm{A}}}
\bmdefine{\bmH}{{\bm{H}}}
\bmdefine{\bmS}{{\bm{S}}}
\newcommand{\Exp}[1]{{\langle\, #1\, \rangle}}
\newcommand{\mfw}{\mathfrak{w}}

\newcommand{\epsmu}{\epsilon_\mu}

\begin{document}


%
\title{
Dynamic critical phenomena in the AdS/CFT duality}
\author{Kengo Maeda}
\email{maeda302@sic.shibaura-it.ac.jp}
\affiliation{Department of Engineering, 
Shibaura Institute of Technology, Saitama, 330-8570, Japan}
\author{Makoto Natsuume}
\email{makoto.natsuume@kek.jp}
\affiliation{Theory Division, Institute of Particle and Nuclear Studies, \\
KEK, High Energy Accelerator Research Organization, Tsukuba, Ibaraki, 305-0801, Japan}
\author{Takashi Okamura}
\email{tokamura@kwansei.ac.jp}
\affiliation{Department of Physics, Kwansei Gakuin University,
Sanda, Hyogo, 669-1337, Japan}
\date{\today}
\begin{abstract}
In critical phenomena, singular behaviors arise not only for thermodynamic quantities but also for transport coefficients. We study this dynamic critical phenomenon in the AdS/CFT duality. We consider black holes with a single R-charge in various dimensions and compute the R-charge diffusion in the linear perturbations. In this case, the black holes belong to model B according to the classification of Hohenberg and Halperin.
\end{abstract}

\maketitle

\section{Introduction}\label{sec:intro}

The AdS/CFT duality is a powerful tool to study gauge theories at strong coupling. At zero temperature, there are many circumstantial evidences for the duality, but the finite temperature, dynamic cases are less understood. In fact, the results for these cases such as the universality of $\bareta/s$ ($\bareta$: shear viscosity,%
\footnote{In this paper, we use $\bareta$ for the shear viscosity to avoid confusion with a static critical exponent $\eta$ (See \sect{static}).} $s$: entropy density) are mostly regarded as predictions for dual gauge theories. (See Ref.~\cite{Natsuume:2007qq} for a review.)
Obviously, the problem comes from the fact that it is extremely hard to compute gauge theories at strong coupling. 

The critical phenomena may be useful to check some aspects of the duality.
At the critical point, the correlation length $\xi$ diverges, and only a few low-energy, long-wavelength modes become dominant. As the consequence, it is possible to capture the behavior of physics which does not depend on the details of microscopic physics. This property enables one to check the AdS/CFT duality even if one cannot compute gauge theories at strong coupling. 

At the equilibrium, the effect of diverging correlation length appears as the divergence of thermodynamic quantities such as the specific heat ({\it static critical phenomena}). The divergence is parametrized by the static critical exponents. For normal statistical systems, the effect of diverging correlation length also appears in nonequilibrium cases ({\it dynamic critical phenomena}). For example, the relaxation time%
\footnote{One should not confuse this relaxation time with the one appeared in the second-order hydrodynamics \cite{causal_review}. }
and the other transport coefficients diverge, which is known as the {\it critical slowing down}. The divergence is parametrized by the dynamic critical exponents, and these dynamic critical exponents are related to the static critical exponents. Our aim is to see these singular behaviors from black hole physics and to see if they agree with the predictions from the theory of dynamic critical phenomena. 

The simplest AdS/CFT duality involves the Schwarzschild-AdS$_5$ (SAdS$_5$) black hole (with planar horizon), which has no phase transition at finite temperature. The SAdS$_5$ black hole with compact horizon such as $S^3$ can have a phase transition, but it is not suitable for our purpose: First, it is a first-order phase transition; Second, the theory has no hydrodynamic limit [$\omega(q) \rightarrow 0$ as $q \rightarrow 0$] since $q$ takes only the discrete values. Similarly, the Reissner-Nordstr\"{o}m-AdS$_5$ black hole with compact horizon has a critical point similar to QCD, but it is not suitable since it has no hydrodynamic limit. A simple system which has both of the second-order phase transition and the hydrodynamic limit is the $D=5$ single R-charge black hole with planar horizon \cite{Behrndt:1998jd,Kraus:1998hv,Cvetic:1999xp}.%
\footnote{In this paper, we denote $D$ as the number of bulk (noncompact) spacetime dimensions and $d$ as the number of boundary spatial dimensions.}

In this paper, we consider single R-charge black holes in various dimensions. 
We consider the linear perturbations in the backgrounds, compute the relaxation time for the R-charge diffusion, and show the critical slowing down. We show that the black holes belong to model B according to the classification of Hohenberg and Halperin \cite{hohenberg_halperin}, {\it i.e.}, the same universality class as the uniaxial ferromagnet.  (One can regard our systems as model H, the same universality class as the liquid-gas phase transition. See \sect{modelH}.)

In the next section, we briefly review both static and dynamic critical phenomena \cite{textbook}. We analyze the dynamic critical phenomena for the $D=5$ single R-charge black hole in \sect{main}. In \sect{mbranes}, we extend the analysis to the $D=4, 7$ single R-charge black holes. We discuss some related issues in \sect{discussion}.%
\footnote{Recently, the dynamic critical exponent has been discussed in the context of the Schr\"{o}dinger group in Refs.~\cite{Balasubramanian:2008dm,Maldacena:2008wh,Adams:2008wt,Minic:2008xa,Kachru:2008yh,Kovtun:2008qy}. 
See also Refs.~\cite{Son:2008ye,Herzog:2008wg} which overlap with these works.
}

\section{Critical phenomena}\label{sec:critical_phenomena}

\subsection{Static critical phenomena}\label{sec:static}

For the second-order phase transition, thermodynamic quantities diverge as powers of the correlation length $\xi$, and these powers, static critical exponents, are universal. Namely, they depend on the symmetry, spatial dimensionality, and so on, but not the other details of the interactions. Different physical systems may belong to the same universality class; {\it e.g.}, the 3d Ising model and the liquid-gas phase transition. These critical exponents also satisfy static scaling relations, which implies that not all critical exponents are independent. In fact, these scaling relations are derived from a scaling law for the thermodynamic potential.

As an example, consider the ferromagnetic phase transition. In this case, the magnetization $m$ and the external magnetic field $h$ are the order parameter and a control parameter, respectively, 
which are related by
\be
m = -\left(\frac{\del \Omega}{\del h}\right)_T~,
\ee
where $\Omega = \Omega(T,h)$ is the Gibbs free energy whose variation is given by
\be
\Omega = - s\, dT - m\, dh~.
\ee
The static critical exponents $(\alpha, \beta, \gamma, \delta, \nu, \eta)$ are defined by
\begin{subequations}
\label{eq:exponents}
\begin{align}
&\mbox{specific heat:} & C_h
&:= - T \left(\frac{\del^2 \Omega}{\del T^2}\right)_h
\propto |\eT|^{-\alpha}~, & \\
&\mbox{spontaneous magnetization:} & m &\propto |\eT|^{\beta} & (T<T_c), \\
&\mbox{magnetic susceptibility:} & \chi_T 
&:= \left(\frac{\del m}{\del h}\right)_T 
\propto |\eT|^{-\gamma}~, & \\
&\mbox{critical isotherm:} & m & \propto |h|^{1/\delta} & (T=T_c), \\
&\mbox{correlation function:} 
&   G(\vec{r}\,)  & \propto e^{-r/\xi} & (T \neq T_c), \\
&&         & \propto r^{-d+2-\eta} & (T=T_c), \\
&\mbox{correlation length:} & \xi  &\propto |\eT|^{-\nu}~,
\label{eq:nu-def} &
\end{align}
\end{subequations}
where $\eT:= (T-T_c)/T_c$, and $d$ denotes the number of {\it spatial} dimensions.
The correlation function is defined by 
\be
G(\vec{r}\,) := \langle  m(\vec{r}\,)m(0) \rangle 
\propto 
\frac{\del^2 \Omega}{\del h(\vec{r}\,) \del h(0)}~.
\ee
By definition, $\chi_T \propto \tilG(\vecq=0)$, where $\tilG(\vecq\,)$ is the Fourier component of $G(\vec{r}\,)$. Since 
$\tilG(q) \propto q^{-2+\eta} \propto \xi^{2-\eta} (\xi q)^{-2+\eta}$, 
the susceptibility diverges at the critical point since the correlation length diverges. In fact, the scaling law below claims that the divergence of all thermodynamic quantities is due to the diverging $\xi$. 

There are 6 static critical exponents in \eq{exponents}, but not all are independent, and they satisfy static scaling relations:
\begin{subequations}
\bea
\alpha + 2\beta+\gamma =2~, 
\label{eq:scaling1} \\
\gamma = \beta(\delta-1)~, 
\label{eq:scaling2} \\
\gamma = \nu (2-\eta)~, 
\label{eq:scaling3} \\
2-\alpha = \nu d~.
\label{eq:scaling4} 
\eea
\end{subequations}
A relation involving $d$ such as \eq{scaling4} is known as a hyperscaling relation. Because of these relations, only 2 are independent among 6 exponents, which suggests that there is some structure behind these relations. 

In fact, these relations can be derived from the static scaling law for the free energy:
\be
\Omega (\eT, h) = b^{-d} \Omega(b^{y_T} \eT, b^{y_h} h)~.
\label{eq:static_law_energy}
\ee
Equation~(\ref{eq:static_law_energy}) implies
\be
\Omega (\eT, h) = \eT^{d/y_T} \Omega(1, \eT^{-y_h/y_T} h)
 =: \eT^{d/y_T} \Psi (\eT^{-y_h/y_T} h)~.
\label{eq:static_law_energy2}
\ee
The scaling law gives a scaling law for the correlation function:
\be
\tilG (\vecq, \eT, h) 
= b^{2y_h-d} \tilG(b\vecq, b^{y_T} \eT, b^{y_h} h)~.
\label{eq:static_law_correlator1}
\ee
The static scaling law determines 6 critical exponents in terms of 2 parameters $(y_T,y_h)$:
\begin{subequations}
\bea
\alpha &=& 2-d/y_T~, \label{eq:alpha} \\
\beta &=& \frac{d-y_h}{y_T}~, \label{eq:beta} \\
\gamma &=& \frac{2y_h-d}{y_T}~, \label{eq:gamma} \\
\delta &=& \frac{y_h}{d-y_h}~, \label{eq:delta} \\
\nu &=& 1/y_T~, \label{eq:nu} \\
\eta &=& d - 2y_h + 2~. \label{eq:eta} 
\eea
\end{subequations}
Equations~(\ref{eq:alpha})-(\ref{eq:eta}) satisfy the scaling relations (\ref{eq:scaling1})-(\ref{eq:scaling4}). It is useful to rewrite the scaling law (\ref{eq:static_law_correlator1}) in a form similar to \eq{static_law_energy2}:
\be
\tilG (\vecq, \xi, h) 
= \xi^{2-\eta} \Phi(\xi \vecq, \xi^{y_h} h)~,
\label{eq:static_law_correlator}
\ee
where Eqs.~(\ref{eq:nu}) and (\ref{eq:eta}) are used. Note $\vec{q}$ appears in the form $(\xi \vec{q})$: this is a typical form in a scaling law. 

As an example, consider the (uniaxial) ferromagnets such as the Ising model. 
In the mean-field approximation, the Ising model is described by the Ginzburg-Landau theory: 
\be
I[m; T, h] = \int d\vecx \,\left[ \frac{c}{2}\,(\nabla m)^2+\frac{a}{2}\,m^2 + \frac{b}{4}\, m^4+\cdots-m h \right]~,
\label{eq:GL_free_energy}
\ee 
where $a=a_0(T-T_c)+\cdots$, $b=b_0+\cdots$, and $c=c_0+\cdots$ (The dots represent the terms which do not contribute near the critical point). 
As usual for the Legendre transformation, $I[m; T, h]$ is a pseudofree energy, and the free energy is given by eliminating $m$: $\Omega(T,h) = \min_m I[m; T, h]$. The static critical exponents for the Ginzburg-Landau theory are 
\be
(\alpha, \beta, \gamma, \delta, \nu, \eta) = \left(0, \frac{1}{2}, 1, 3,  \frac{1}{2}, 0\right)~
\label{eq:static_exponent_LG}
\ee 
for $d=3$. These exponents satisfy the scaling relations except the hyperscaling relation (\ref{eq:scaling4}). In fact, it is easy to show that the pseudofree energy (\ref{eq:GL_free_energy}) satisfies the scaling law (\ref{eq:static_law_energy}). 

There are some caveats which are relevant to our later discussion: 
\begin{itemize}
\item  
The mean-field theory ignores the effects of statistical fluctuations. 
The Ginzburg criterion tells that the mean-field theory is trustable for $d>d_c$, where $d_c$ is the upper critical dimension given by
\be
d_c = \frac{2\beta+\gamma}{\nu} = \frac{2-\alpha}{\nu}~.
\label{eq:critical_dim}
\ee
Namely, the static critical exponents obtained from a mean-field theory is unreliable {\it below} $d_c$. For the Ginzburg-Landau theory, $d_c=4$. On the other hand, we consider the large-$N_c$ limit, where the effects of fluctuations are suppressed. 

\item 
The hyperscaling relation (\ref{eq:scaling4}) may fail {\it above} $d_c$ (or even below it in some problems). This is because the scaling law for the free energy (\ref{eq:static_law_energy}) is modified by the so-called dangerous irrelevant operators.
\end{itemize}

\subsection{Dynamic critical phenomena}\label{sec:dynamic}

In the dynamic case, the order parameter varies slowly in time. Near the critical point, a large domain is favorable due to the diverging correlation length. Because of the large size, it takes a long time for thermal fluctuations to flip spins for a whole domain. This is known as the critical slowing down, and the divergence of the relaxation time is parametrized by the dynamic critical exponent $z$.

The idea of scaling law can be naturally extended into the dynamic case. The static scaling law for the correlation function (\ref{eq:static_law_correlator}) is extended into the dynamic scaling law:
\bea
\tilG (\omega, \vecq, \eT, h) 
&=& b^{2-\eta+z} \tilG(b^z \omega, b\vecq, b^{y_T} \eT, b^{y_h} h) \\
&=& \xi^{2-\eta+z} \Phi(\xi^z \omega, \xi \vecq, \xi^{y_h} h)~,
\eea
where $z$ is a dynamic critical exponent. The scaling law suggests that the relaxation time diverges near the critical point:
\begin{subequations}
\label{eq:slowing_from_scaling}
\begin{alignat}{3}
   \tau_q \sim \omega_{\rm typical}^{-1} 
  &\sim \xi^z \varphi(\xi q)
& \hspace{0.2cm}
  & \rightarrow \infty
& \hspace{2.0cm}
  & (T \rightarrow T_c)~,
\label{eq:slowing_from_scaling1} \\
  &\sim q^{-z}(\xi q)^z \varphi(\xi q)
& & \propto q^{-z}
& & (T=T_c)~.
\label{eq:slowing_from_scaling2}
\end{alignat}
\end{subequations}
[In reality, the fluctuation-dissipation theorem is necessary to show Eqs.~(\ref{eq:slowing_from_scaling}), and we skip some steps here.]

The dynamic universality classes were classified by Hohenberg and Halperin \cite{hohenberg_halperin}: they are known as model A, B, C, H, F, G, and J. These models are further classified by the values of dynamic critical exponents. 
The dynamic universality class depends on additional properties of the system which do not affect the static universality class. In particular, conservation laws play an important role to determine dynamic critical exponents. A conservation law forces the relaxation to proceed more slowly. 
As a consequence, even if two systems belong to the same static universality class, they may not belong to the same dynamic universality class. 

We consider model B example below. (See \sect{modelH} for model H.) 
Model B corresponds to a system with a conserved charge. This is the case for uniaxial ferromagnets, but we use the conventions for the charge diffusion below for a later purpose. In the hydrodynamic limit, it is governed by the diffusion equation:
\begin{align}
  & 0 = \partial_t \rho(t, \vecx\,) - D_R\, \Delta \rho(t, \vecx\,)
& & \Longleftrightarrow
& & 0 = \left( \omega + i\, D_R\, q^2 \right)\, \rho(\omega, \vecq\,)~,
\label{eq:R_charge-diffusion_eqn}
\end{align}
where $D_R$ is the diffusion constant. 
Note that the diffusion constant $D_R$ is related to the susceptibility $\chi_T$ and the conductivity $\lambda$ as 
\be
\lambda = D_R \chi_T~,
\label{eq:conductivity}
\ee
from the definitions of $\lambda$ and $D_R$:
\bea
J_i &=& -\lambda \nabla_i\mu 
= -\lambda \left(\frac{\partial \mu}{\partial \rho} \right)_{T}\nabla_i \rho~,
\\
J_i &=& -D_R \nabla_i \rho~.
\eea

It is important to distinguish three different regimes: the hydrodynamic regime  ($\xi q \ll 1$), the critical regime  ($\xi q \gg 1$), and the overlapped regime ($\xi q \approx 1$).

\begin{enumerate} 
\item The hydrodynamic regime: 
In this regime, the charge diffusion is governed by the diffusion equation, 
and the relaxation time $\tau_q$ for wave number  $q$ is given by 
\be
1/\tau_q \sim - \Im(\omega) = D_R q^2~.
\label{eq:hydro_relaxation}
\ee
{\it For model B, $\lambda$ is nonsingular on the critical point.} Because $\chi_T$ diverges on the critical point, this means that $D_R$ vanishes on the critical point, namely the critical slowing down. 
Then, Eqs.~(\ref{eq:conductivity}) and (\ref{eq:hydro_relaxation}) give 
\be
1/\tau_q \propto \chi_T^{-1}\, q^2~.
  \label{eq:dynamic_critical_exp2} 
\ee

\item The critical regime: On the other hand, in the critical regime, higher powers of $q$ in the dispersion relation are no longer negligible. In this regime, the dynamic scaling relation (\ref{eq:slowing_from_scaling}) dictates the dispersion relation $1/\tau_q \propto q^z$. 

\item The overlapped regime: These two behaviors should match smoothly at $\xi q \approx 1$.
Thus, \eq{dynamic_critical_exp2} should be rewritten as 
\be
1/\tau_q \propto \xi^{-(4-\eta)}\, (\xi q)^2~,
\label{eq:dynamic_critical_exp3}
\ee
where a static scaling relation is used.
\end{enumerate}
Comparing \eq{dynamic_critical_exp3} with \eq{slowing_from_scaling1}, one gets 
\be
z = 4 - \eta~.
\label{eq:dynamic_static_law}
\ee
Note that \eq{dynamic_static_law} relates the dynamic critical exponent $z$ to the static critical exponent $\eta$.  

The important assumption of model B is that $\lambda$ remains finite on the critical point. We will check that this is indeed the case for single R-charge black holes.

\section{Dynamic critical phenomena for the $D=5$ single R-charge black hole}\label{sec:main}

In this section, we show that the critical slowing down indeed occurs at the second-order phase transition for the $D=5$ single R-charge black hole \cite{Behrndt:1998jd}.

\subsection{The $D=5$ single R-charge black hole and its thermodynamics}

In the simplest AdS/CFT duality, the gravity dual is SAdS$_5 \times S^5$. One can add an angular momentum along $S^5$, which is known as the ``spinning" D3-brane solutions \cite{Kraus:1998hv,Cvetic:1999xp}. The angular momentum becomes a charge after the $S^5$ reduction. The symmetry of $S^5$ corresponds to an internal symmetry of the ${\cal N}=4$ super-Yang-Mills theory, R-symmetry $SU(4)_R$. The R-symmetry group $SU(4)_R$ is rank 3, so one can add at most three independent charges. The three-charge solution is known as the STU solution \cite{Behrndt:1998jd}. When all charges are equal, the STU solution is nothing but the Reissner-Nordstr\"{o}m-AdS$_5$ black hole.

The $S^5$ reduction of $D=10$ type IIB supergravity gives $D=5$ $\mathcal{N}=2$ gauged U(1)$^3$ supergravity \cite{Cvetic:1999xp}:
%
\begin{align}
   \calL_5
  &= \sqrt{-g}~\left[~R
  - \frac{1}{4}\, G_{ij}\, F^i_{\mu\nu}\, F^{j\,\mu\nu}
  - \frac{1}{2}\, G_{ij}\, \left( \nabla_\mu X^i \right)\,
                           \left( \nabla^\mu X^j \right)
  + \frac{4}{L^2}\, \sum_{i=1}^3 \frac{1}{X^i}~\right]
\nonumber \\
  &\hspace*{2.0cm}
  + \frac{1}{24}\, \epsilon^{\mu\nu\rho\sigma\lambda}\,\epsilon_{ijk}\,
     F^i_{\mu\nu} F^j_{\rho\sigma} A^k_{\lambda}~.
\label{eq:reduced-action-3charge}
\end{align}
Here, we use the unit $2\, \kappa_5^2 = 16 \pi G_5 = 1$, and 
$F^i_{\mu\nu}$ are the field strength of the 3 U(1) gauge fields
$A^i_\mu$~($i= 1, 2, 3$).
The fields $X^i$~($i=1, 2, 3$) represent 3 real scalar fields which are not independent but are subject to the constraint 
$X^1\, X^2\, X^3 = 1$,  and their moduli space metric $G_{ij}$ is given by
\be
G_{ij} = \text{diag}[~(X^1)^{-2},\, (X^2)^{-2},\, (X^3)^{-2}~]~.
\ee

For the single R-charge black hole, use the ansatz
\begin{align}
  & F_{\mu\nu}^2 = F_{\mu\nu}^3 = 0~,
& & X^2 = X^3 = 1/\sqrt{X^1} =: H^{1/3}~,
\end{align}
which is consistent with the equations of motion.
Then, the action becomes 
\begin{align}
   \calL_5
  &= \sqrt{-g}\, \left[~R - \frac{L^2}{8}\, H^{4/3}\, F^2
  - \frac{1}{3}\, \frac{(\nabla H)^2}{H^2}
  + 2 {\cal V}~\right]~,
\label{eq:reduced-action-1charge}
\end{align}
where 
\be
2{\cal V} = \frac{4}{L^2}\, \left( H^{2/3} + 2 H^{-1/3} \right)~,
\ee
$A_\mu := (\sqrt{2}/L)\, A^1_\mu$, and $F_{\mu\nu} := 2 \partial_{[\mu} A_{\nu]}$. 
The equations of motion are given by
\begin{subequations}
\begin{align}
  & R_{\mu\nu}
  = 3\, \frac{(\nabla_\mu H)(\nabla_\nu H)}{H^2}
  + \frac{L^2}{4}\, H^{4/3}\, F_{\mu\lambda} F_{\nu}{}^{\lambda}
  - \frac{g_{\mu\nu}}{6}\,
    \left( \frac{L^2}{4}\, H^{4/3}\, F^2 + 4\, \calV \right)~,
\label{eq:EOM_grav} \\
  & \nabla_\nu \left( H^{4/3}\, F^{\mu\nu} \right) = 0~,
\label{eq:EOM_EM} \\
  & \square \ln H
  = - 3\, H\, \frac{d\calV}{dH}
  + \frac{L^2}{4}\, H^{4/3}\, F^2~.
\label{eq:EOM_scalar-II}
\end{align}
\end{subequations}
%
%
and the single R-charge black hole (with planar horizon) is given by
\begin{subequations}
\begin{align}
  ds_5^2
  &= \frac{(\pi\, T_0\, L)^2}{u}\, H^{1/3}\,
    \left( - \frac{f}{H}\, dt^2 + dx^2 + dy^2 + dz^2 \right)
  + \frac{L^2}{4\, f\, u^2}\, H^{1/3}~du^2~,
\label{eq:STU_metric-five_dim-1charge} \\
  A_{\mu}
  &= \pi\, T_0\, \sqrt{2\, \kappa\, (1+\kappa)}~\frac{u}{H}~( dt )_\mu~,
\label{eq:STU_U1-1charge} \\
  H &= 1 + \kappa\, u~,
\\
  f &= H - (1+\kappa)\, u^2
  = (1-u) \left\{ 1 + (1+\kappa)\, u \right\}~,
\label{eq:STU_scalar-1charge}
\end{align}
\end{subequations}
where $u \in [0,1]$, with $u=1$ corresponding to the location of the horizon.

The black hole is characterized by two parameters
$(T_0, \kappa)$: the former gives the Hawking temperature for the neutral black hole, and the latter gives the R-charge. Thermodynamic quantities for the black hole are given by
\begin{subequations}
\label{eq:thermodynamic_quantities}
\begin{align}
  & \epsilon
  = \frac{3\, \pi^2\, N_c^2\, T_0^4}{8}~( 1 + \kappa )~,
& & P
  = \frac{\pi^2\, N_c^2\, T_0^4}{8}~( 1 + \kappa )~,
\label{eq:energy_density-P-1charge} \\
  & s
  = \frac{\pi^2\,N_c^2\,T_0^3}{2}~\sqrt{1 + \kappa}~,
& & T
  = \frac{2 + \kappa}{2\, \sqrt{1 + \kappa}}~T_0~,
\label{eq:entropy_density-T_H-1charge} \\
  & \rho
  = \frac{\pi\, N_c^2\, T_0^3}{8}~\sqrt{2\, \kappa\, ( 1 + \kappa )}~,
& & \mu
  = \pi\, T_0\, \sqrt{ \frac{2\, \kappa}{1 + \kappa} }~,
\label{eq:charge_density-eq:chemical_potential-1charge}
\end{align}
\end{subequations}
which satisfy the thermodynamic relations:
\begin{align}
  & d\epsilon = T\, ds + \mu\, d\rho~,
& & dP = s\, dT + \rho\, d\mu~,
& & \epsilon + P = s\, T + \mu\, \rho~.
\label{eq:thermodynamic_relation}
\end{align}

Hereafter, we study the black hole as the grand canonical ensemble, {\it i.e.}, the system is characterized by $(T, \mu)$ not by $(T_0, \kappa)$.
The appropriate thermodynamic potential is the Gibbs free energy $\Omega=-P= \epsilon-Ts-\mu\rho$. 
The quantities $T$ and $\mu$ satisfy
\begin{align}
  & \frac{2\, \pi\, T}{\mu}
  = \sqrt{ \frac{\kappa}{2} } + \sqrt{ \frac{2}{\kappa} } \ge 2
& & (\text{The equality holds for $\kappa = 2$})~.
\label{eq:sol-degenerateI}
\end{align}
For a given pair of $(T, \mu)$, there are two black hole solutions. These solutions are related by the transformation
%
\begin{align}
  & \kappa \longrightarrow \frac{4}{\kappa}~,
& & T_0 \longrightarrow T_0\, \sqrt{
     \frac{\kappa\, (4+\kappa)}{4\, (1+\kappa)} }~.
\label{eq:sol-degenerateII}
\end{align}
(This transformation is valid for $D=4, 7$ single R-charge black holes in \sect{mbranes} as well.) The solution thermodynamically realized is the one with the lower Gibbs free energy, which is the branch of $\kappa \le 2$. Thus, we focus on $\kappa \le 2$.

Thermodynamic stability is determined from the behavior of thermodynamic fluctuations, {\it e.g.}, the R-charge susceptibility and the specific heat at constant $\mu$:
\begin{align}
   \chi_T
  &:= \left( \frac{\partial \rho}{\partial \mu} \right)_{T}
  = \frac{N_c^2 T_0^2}{8}~
     \frac{2 + 5 \kappa - \kappa^2}{2 - \kappa}~,
\label{eq:electric_susceptibility-1charge} \\
   C_\mu
  &:= T\, \left( \frac{\partial s}{\partial T} \right)_{\mu}
  = \frac{1}{2}\, \pi^2\, N_c^2 T_0^3~
     \frac{ ( 3 - \kappa ) ( 2 + \kappa ) }
          { \sqrt{1 + \kappa\, }\, ( 2 - \kappa ) }~,
\label{eq:heat_capacity-grand_canonical-1charge}
\end{align}
which diverge at $\kappa = 2$. 
On the other hand, the first derivatives of the Gibbs free energy such as $s$ and $\rho$ are regular there, so this is a second-order phase transition. Note that $\kappa=2$ imposes one condition in the $(T, \mu)$-phase diagram, so $\kappa=2$ defines a critical line instead of a critical point. 

The thermodynamic quantities (\ref{eq:thermodynamic_quantities}) determine the static critical exponents \cite{Cai:1998ji,Cvetic:1999rb}:
\be
(\alpha, \beta, \gamma, \delta, \nu, \eta) = \left(\frac{1}{2}, \frac{1}{2}, \frac{1}{2}, 2,  \frac{1}{2}, 1\right)~.
\label{eq:5d_static_exponents}
\ee
 (See \appen{staticR}.)

\subsection{Critical slowing down for the $D=5$ single R-charge black hole}

The single R-charge black hole has a second-order phase transition at $\kappa=2$ as we saw in the last subsection, so one expects a critical slowing down, where a relaxation time diverges. According to the discussion in \sect{dynamic}, if the single R-charge black hole belongs to model B, $\lambda$ remains finite on the critical line. In this subsection, we consider the linear perturbations for the black hole, and we explicitly show that $\lambda$ remains finite on the critical line. 

The R-charge conductivity $\lambda$ can be evaluated from a Green-Kubo formula:
\begin{align}
  \lambda
  &= - \lim_{\omega \to 0}~
      \frac{\Im\left[~\tilG^{(R)}_{xx}(\omega, \vec{q} = 0)~\right]}
           {\omega} 
\nonumber \\
  &= \lim_{\omega \rightarrow 0}\, \frac{1}{2\, \omega}
  \int^\infty_{-\infty} dt~e^{i \omega t} \int d\vecx~
  \Exp{ [\, J_x(t, \vecx\,)\,,~J_x(0, \Vec{0}\,)\, ] }~,
\label{eq:Kubo_formula}
\end{align}
where $\tilG^{(R)}_{xx}(\omega, \vecq\,)$ is the retarded correlator of the R-charge current:
\begin{align}
   \tilG^{(R)}_{xx}(\omega, \vecq\,)
  &:= - i \int^\infty_{-\infty} dt~e^{i \omega t}~\theta(t)
  \int d\vecx~e^{- i \vecq \cdot \vecx}~
  \Exp{ [\, J_x(t, \vecx\,)\,,~J_x(0, \Vec{0}\,)\, ] }~.
\label{eq:def-tilde_Gxx}
\end{align}
We evaluate this correlator from the AdS/CFT duality.%
\footnote{Our results in this section can be obtained from the results of Ref.~\cite{Son:2006em} (See Sec.~4 of the paper. See also Refs.~\cite{Mas:2006dy,Saremi:2006ep,Maeda:2006by} for related issues), but the computation presented here is easier than their computation. Both our computation and their computation consider the vector modes. They have done the computation to obtain the shear viscosity $\bareta$ whereas we are interested in $D_R$. In the vector modes, the quasinormal mode method is used to obtain $\bareta$ whereas the Green-Kubo formula method is used to obtain $D_R$. As the consequence, the former needs the $q \neq 0$ correlator, and the latter needs only the $q=0$ correlator. Our result is a limit of their result, but the computation is far easier.}
Since \eq{def-tilde_Gxx} is a vector mode correlator, it is enough to consider the vector modes of the bulk fields.%
\footnote{The electromagnetic perturbation or the R-charge current is decomposed as the vector mode and the scalar mode. Similarly, the gravitational perturbation or the energy-momentum tensor is decomposed as the tensor mode, the vector mode, and the scalar mode. }
The perturbations from the background fields $\bmg_{\mu\nu}$, $\bmA_{\mu}$ are denoted as
\begin{align}
  & h_{\mu\nu} := g_{\mu\nu} - \bmg_{\mu\nu}~,
& & a_{\mu} := A_{\mu} - \bmA_{\mu}~.
\end{align}
Since we consider the vector modes, the scalar field $H$ does not fluctuate, {\it i.e.}, $H=\bm{H}$. Thus, we consider the perturbations which take the form 
\begin{align}
  & h_{t x} =: \bmg_{xx}\, T(u)\, e^{- i \omega t + i q z}~,
& & h_{z x} =: \bmg_{xx}\, Z(u)\, e^{- i \omega t + i q z}~,
& & a_{x} =: \frac{\mu}{2}\, A(u)\, e^{- i \omega t + i q z}~,
\end{align}
with the other $h_{\mu\nu}=0$.

Moreover, we need only the $q=0$ correlator since we evaluate the Green-Kubo formula (\ref{eq:Kubo_formula}). This gives a further simplification, namely it is enough to consider the $q = 0$ limit of the equations of motion for $(T,Z,A)$:
\begin{subequations}
\begin{align}
  & Z'' + \frac{u\, f' - f}{u\, f}\, Z'
  + \frac{\mfw^2\, \bmH}{u\, f^2}\, Z
  = 0~,
\label{eq:Eq_Z} \\
  & T' + \frac{\kappa\, u}{2\, \bmH}\, A = 0~,
\label{eq:const} \\
  & A'' + \left( \frac{\bmH'}{\bmH} + \frac{f'}{f}
          \right)^{\mathstrut}\, A'
  + \frac{\mfw^2\, \bmH}{u\, f^2}\, A
  + 2\, \frac{1 + \kappa}{f\, \bmH}\, T' = 0~,
\label{eq:Eq_A}
\end{align}
\end{subequations}
where
\be
  \mfw := \frac{\omega}{2 \pi T_0}~.
\ee
In the $q=0$ limit, the field $Z$ decouples from $A$ and $T$, and Eqs.~(\ref{eq:const}) and (\ref{eq:Eq_A}) give an equation for $A$ only:
\begin{align}
  & A'' + \left( \frac{\bmH'}{\bmH} + \frac{f'}{f}
          \right)^{\mathstrut}\, A'
  + \left( \frac{\mfw^2\, \bmH}{u\, f^2}
  - \frac{\kappa\, (1 + \kappa)}{f\, \bmH^2}\, u \right)\, A
  = 0~.
\label{eq:Eq_A-closed}
\end{align}
Solving this equation is essentially enough to get the R-charge current correlator.

The correlator is obtained from the Lorentzian prescription of the AdS/CFT duality \cite{Policastro:2002se}:
\begin{itemize}
\item Step~1: Solve the bulk equations of motion by imposing the ``incoming wave" boundary condition at the horizon and by imposing the boundary values of the bulk fields. The incoming wave boundary condition corresponds to the retarded condition, and the boundary values play the role of source terms for the boundary fields in the dual gauge theory.

\item Step~2: The total action consists of the bulk action (\ref{eq:reduced-action-1charge}), the Gibbons-Hawking term, and conterterm action which cancels divergences.  Substitute the solution obtained in Step~1 into the action, and evaluate the boundary action quadratic in the boundary values. 

\item Step~3: The correlator is obtained from the second derivative of the boundary action with respect to the boundary values.
\end{itemize}
The first is to solve the bulk equation of motion (\ref{eq:Eq_A-closed}), which can be done perturbatively in $\mfw$. Incorporating the incoming wave boundary condition at the horizon, we get 
\begin{align}
  & a_x(u)
  = a_x^{(0)}\, \frac{ f^{-\frac{i \mfw T_0}{2T}} }{H}~
  \left\{ 1 + \frac{\kappa\, u}{2}
  + \frac{i\, \mfw}{4}~\left(
  \frac{\kappa^2\, u}{\sqrt{1+\kappa}}
    + \frac{2\, T_0}{T}\, (2+\kappa\, u) \ln(u+H) \right)~\right\}
  + O(\omega^2)~,
\label{eq:pert-A_sol}
\end{align}
where $a_x^{(0)}$ represents the boundary value of $a_x(u)$:
\begin{align}
  & \lim_{u \to 0}~a_x(u) = a_x^{(0)}~.
\end{align}
The second is to obtain the boundary action. The boundary action is given by
\begin{align}
   S_{\text{boundary}}
  &= \frac{N_c^2\, T_0^2}{16}~\lim_{u \to 0} \int dt\, d\vecx~
    \left( f\, H\, a_x'\, a_x + \cdots \right)
\nonumber \\
  &= - \frac{N_c^2\, T_0^2}{32}~a_x^{(0)}\, \left( \kappa
     - i\, \omega\, \frac{ (\kappa + 2)^2 }
              { 4\, \pi\, T_0\, \sqrt{1 + \kappa} } \right)~a_x^{(0)}~,
\label{eq:on_shell-action}
\end{align}
where the dots represent the terms which are not quadratic in $a_x^{(0)}$.
Finally, following Step~3, we get
\begin{align}
  & \tilG^{(R)}_{xx}(\omega, q = 0)
  = \frac{\kappa\, N_c^2\, T_0^2}{16}
  - i\, \omega\, \frac{ (\kappa + 2)^2\, N_c^2\, T_0}
           { 64\, \pi\, \sqrt{1 + \kappa} }~,
\label{eq:Rcharge-current_correlator-1charge}
\end{align}
or
\be
  \lambda
  = - \lim_{\omega \to 0}~
      \frac{\Im\left[~\tilG^{(R)}_{xx}(\omega, q = 0)~\right]}
           {\omega}
  = \frac{ (\kappa + 2)^2\, N_c^2\, T_0 }
         { 64\, \pi\, \sqrt{1 + \kappa} }~,
\label{eq:Im_Rcharge-current_correlator-1charge}
\ee
so $\lambda$ is finite on the critical line $\kappa = 2$ as promised.

Using Eq.~(\ref{eq:electric_susceptibility-1charge}), the R-charge diffusion constant is given by
\begin{align}
  & D_R
  = \frac{1}{2\, \pi\, T}~\frac{ (1 + \kappa/2)^3 }{1 + \kappa}~
     \frac{2 - \kappa}{2 + 5\, \kappa - \kappa^2}~,
\label{eq:D_R-1charge}
\end{align}
so $D_R = 0$ on the critical line. 

To summarize, the dynamic universality class for the single R-charge black hole is model B with dynamic critical exponent $z = 4-\eta$ as predicted from the theory of dynamic critical phenomena.

\section{Single R-charge black holes in other dimensions}\label{sec:mbranes}

The extension to other dimensions is straightforward and similar. We consider the $D=4, 7$ cases, which are obtained from rotating M2 and M5-branes, respectively. For the actions  and the solutions, see, {\it e.g.}, Ref.~\cite{Cvetic:1999xp}.

\subsection{$D=4$ single R-charge black hole}

For a single R-charge black hole, the $D=4$ action is given by
\be
   \calL_4
  = \sqrt{-g} \left[~R - \frac{L^2}{8} H^{3/2} F^2
  - \frac{3}{8} \frac{(\nabla H)^2}{H^2}
  + 2{\cal V}~\right]~,
\label{eq:action_4d}
\ee
where
\be
2{\cal V} = \frac{3}{L^2}(H^{1/2}+H^{-1/2})~.
\ee
The single R-charge solution is
\begin{subequations}
\bea
ds_4^2
  &=& \frac{16(\pi T_0 L)^2}{9u^2} H^{1/2}
    \left( - \frac{f}{H} dt^2 + dx^2 + dz^2 \right)
  + \frac{L^2}{f u^2} H^{1/2}~du^2~, \\ 
A_{\mu}  &=& \frac{4}{3} \pi T_0 \sqrt{2 \kappa(1+\kappa)}~\frac{u}{H}~( dt )_\mu~, \\  
H &=& 1 + \kappa\, u~, \\
f &=& (1-u) \left\{ 1 + (1+\kappa) u +(1+\kappa)u^2 \right\}~,
\eea
\end{subequations}
which satisfies the equations of motion from \eq{action_4d}.

Thermodynamic quantities are given by 
\begin{subequations}
\begin{align}
  & \epsilon
  = \sqrt{2}\, \pi^2\, \left( \frac{2}{3} \right)^4\,
    N_c^{3/2}\, T_0^3~(1 + \kappa)~,
& & P
  = \frac{\sqrt{2}\, \pi^2}{3}\, \left( \frac{2}{3} \right)^3\,
    N_c^{3/2}\, T_0^3~(1 + \kappa)~,
\label{eq:energy_density-P-M2} \\
  & s  = \sqrt{2}\, \pi^2\, \left( \frac{2}{3} \right)^3\,
    N_c^{3/2}\, T_0^2~\sqrt{1 + \kappa}~,
& & T
  = \frac{T_0}{\sqrt{1 + \kappa}}~
  \left( 1 + \frac{2\, \kappa}{3} \right)~,
\label{eq:entropy_density-T_H-M2} \\
  & \rho
  = \sqrt{2}\, \pi\, \left( \frac{1}{3} \right)^3\,
    N_c^{3/2}\, T_0^2~\sqrt{2\, \kappa\, ( 1 + \kappa )}~,
& & \mu
  = \frac{4 \pi\, T_0}{3}\, \sqrt{ \frac{2\, \kappa}{1 + \kappa} }~,
\label{eq:charge_density-chemical_potential-M2}
\end{align}
\end{subequations}
from which one gets
\begin{align}
  & \frac{\pi\, T}{\mu}
  = \frac{\sqrt{3}}{4} \left(\sqrt{ \frac{3\kappa}{2} } + \sqrt{ \frac{2}{3\kappa} } \right)
     \ge \frac{\sqrt{3}}{2}
& & (\text{The equality holds for $\kappa = 3/2$})~.
\label{eq:4dsol-degenerate}
\end{align}
Again there are two solutions of $\kappa$ for a single pair of $(T, \mu)$, and we focus on $\kappa \le 3/2$. 
The R-charge susceptibility and the specific heat at constant $\mu$ are given by
\begin{align}
   \chi_T
   &= \frac{N_c^{3/2}\, T_0}{6 \sqrt{2}}
     \frac{1 + 2\, \kappa}{3 - 2\, \kappa}~,
\label{eq:electric_susceptibility-M2} \\
   C_\mu
   &= \sqrt{2}\, \pi^2\, \left( \frac{2}{3} \right)^3\, N_c^{3/2}\, T_0^2~
     \frac{ ( 2 - \kappa ) ( 3 + 2\, \kappa ) }
          { \sqrt{1 + \kappa\, }\, ( 3 - 2\, \kappa ) }~.
\label{eq:heat_capacity-grand_canonical-M2}
\end{align}
Thus, $\kappa=3/2$ is the critical line. The static critical exponents $(\alpha,\beta,\gamma,\delta)$ are the same as the $D=5$ case, and $(\nu, \eta)=(3/4,4/3)$ from \eq{static_scaling_exp-III_generic} \cite{Cai:1998ji,Cvetic:1999rb}. 

Similar to the $D=5$ case, the equations of motion for $(T,Z,A)$ in the $q = 0$ limit are given by
\begin{subequations}
\label{eq:4dEOMs}
\begin{align}
  & Z'' + \frac{u\, f' - 2f}{u\, f}\, Z'
  + \frac{9\mfw^2\, \bmH}{4\, f^2}\, Z
  = 0~,
\label{eq:4dEq_Z} \\
  & T' + \frac{\kappa\, u^2}{2\, \bmH}\, A = 0~,
\label{eq:4dconst} \\
  & A'' + \left( \frac{\bmH'}{\bmH} + \frac{f'}{f}
          \right)^{\mathstrut}\, A'
  + \frac{9\mfw^2\, \bmH}{4f^2}\, A
  + \frac{2(1 + \kappa)}{f\, \bmH}\, T' = 0~.
\label{eq:4dEq_A}
\end{align}
\end{subequations}
Equations~(\ref{eq:4dEOMs}) give an equation for $A$ only:
\begin{align}
  & A'' + \left( \frac{\bmH'}{\bmH} + \frac{f'}{f}
          \right)^{\mathstrut}\, A'
  + \left( \frac{9\mfw^2\, \bmH}{4f^2}
  - \frac{\kappa\, (1 + \kappa)}{f\, \bmH^2}\,u^2 \right)\, A
  = 0~.
\label{eq:4dEq_A-closed}
\end{align}
Then, the solution is given by
\begin{align}
   A
   &=: a_x^{(0)}\, \frac{f^{-i \mfw T_0/2 T}}{H}\, \calA + O(\omega^2)~, \\
   \calA
  &= 1 + \frac{2 \kappa u}{3}
\nonumber \\
  &- i\, \mfw\,
  \Bigg[~\frac{2 \kappa^3 ~u}{3 (3 - \kappa) \sqrt{1 + \kappa} }
  - \frac{3 + 2 \kappa\, u}{4 (3 - \kappa)^2 (3 + 2 \kappa)}
  \Bigg\{ 3 \sqrt{\kappa +1}\, (3 - \kappa)^2\, \ln\frac{f}{1 - u}
\nonumber \\
  &\hspace*{2.0cm}
  + 2 \sqrt{3-\kappa } \big\{ \kappa (\kappa + 12) + 9 \big\}
    \arctan\frac{ \sqrt{(3-\kappa)(1+\kappa)}~u }
                  { 2 + (1 + \kappa) u } \Bigg\}~\Bigg]~.
\label{eq:sol_calA-M2}
\end{align}

The boundary action is given by
\begin{align}
   S_{\text{boundary}}
  &= \frac{N_c^{3/2}\, T_0}{36\sqrt{2}}~\lim_{u \to 0} \int dt\, d\vecx~
    \left( f\, H\, a_x'\, a_x + \cdots \right)~.
\end{align}
Substituting \eq{sol_calA-M2} into the boundary action, we get the R-charge current correlator $\tilG^{(R)}_{xx}(\omega, q = 0)$, the R-charge conductivity $\lambda$, and the R-charge diffusion constant $D_R$:
\begin{align}
  & \tilG^{(R)}_{xx}(\omega, q = 0) = \frac{\kappa\, N_c^{3/2}\, T_0}{54 \sqrt{2}}
  - i\, \omega\, \frac{ (3 + 2 \kappa)^2\, N_c^{3/2} }
                      {6^3 \pi\, \sqrt{2 (1 + \kappa)}}~, \\
  & \lambda
  = \frac{(3+2\kappa)^2 N_c^{3/2}}
         { 6^3 \pi\, \sqrt{2(1 + \kappa)} }~, \\
  & D_R
  = \frac{3}{4 \pi T}~\frac{ (1 + 2\kappa/3)^3 (1 - 2\kappa/3) }{(1 + \kappa)(1+ 2\kappa) }~.
\end{align}
Thus, $\lambda$ is finite and $D_R=0$ on the critical line $\kappa=3/2$.

\subsection{$D=7$ single R-charge black hole}

For a single R-charge black hole, the $D=7$ action is given by
\be
   \calL_7
  = \sqrt{-g} \left[~R - \frac{L^2}{8} H^{6/5} F^2
  - \frac{3}{10} \frac{(\nabla H)^2}{H^2}
  + 2{\cal V}~\right]~,
\label{eq:action_7d}
\ee
where
\be
2{\cal V} = \frac{4}{L^2} \left( \frac{3}{2} H^{4/5}+6H^{-1/5} \right)~.
\ee
The single R-charge solution is
\begin{subequations}
\bea
ds_7^2
  &=& \frac{4(\pi T_0 L)^2}{9u} H^{1/5}
    \left( - \frac{f}{H} dt^2 + dx_1^2 + \cdots + dx_4^2 + dz^2 \right)
  + \frac{L^2}{4 f u^2} H^{1/5}~du^2~, \\ 
A_{\mu}  &=& \frac{2}{3} \pi T_0 \sqrt{2 \kappa(1+\kappa)}~\frac{u^2}{H}~( dt )_\mu~, \\  
H &=& 1 + \kappa u^2~, \\
f &=& (1-u) \left\{ 1 + u +(1+\kappa)u^2 \right\}~,
\eea
\end{subequations}
which satisfies the equations of motion from \eq{action_7d}.

Thermodynamic quantities are given by 
\begin{subequations}
\begin{align}
  & \epsilon
  = \frac{5\, \pi^3}{2}\, \left( \frac{2}{3} \right)^7\,
    N_c^3\, T_0^6~( 1 + \kappa )~,
& & P
  = \frac{\pi^3}{2}\, \left( \frac{2}{3} \right)^7\,
    N_c^3\, T_0^6~( 1 + \kappa )~,
\label{eq:energy_density-P-M5} \\
  & s  = 3\, \pi^3\, \left( \frac{2}{3} \right)^7\,
    N_c^3\, T_0^5~\sqrt{1 + \kappa}~,
& & T
  = \frac{T_0}{\sqrt{1 + \kappa}}~
  \left( 1 + \frac{\kappa}{3} \right)~,
\label{eq:entropy_density-T_H-M5} \\
  & \rho
  = \pi^2\, \left( \frac{2}{3} \right)^6\,
    N_c^3\, T_0^5~\sqrt{2\, \kappa\, ( 1 + \kappa )}~,
& & \mu
  = \frac{2 \pi\, T_0}{3}\, \sqrt{ \frac{2\, \kappa}{1 + \kappa} }~,
\label{eq:charge_density-chemical_potential-M5}
\end{align}
\end{subequations}
from which one gets
\begin{align}
  & \frac{\pi\, T}{\mu}
  = \frac{1}{2}\sqrt{\frac{3}{2}} \left(\sqrt{ \frac{\kappa}{3} } + \sqrt{ \frac{3}{\kappa} } \right)
     \ge \sqrt{\frac{3}{2}}
& & (\text{The equality holds for $\kappa = 3$})~.
\label{eq:7dsol-degenerate}
\end{align}
Again there are two solutions of $\kappa$ for a single pair of $(T, \mu)$, and we focus on $\kappa \le 3$. 
The R-charge susceptibility and the specific heat at constant $\mu$ are given by
\begin{align}
   \chi_T
    &= 2 \pi\, \left( \frac{2}{3} \right)^4\,
    N_c^3\, T_0^4~\frac{1 + 4\, \kappa - \kappa^2}{3 - \kappa}~,
\label{eq:electric_susceptibility-M5} \\
   C_\mu
  &= 2\, \pi^3\, \left( \frac{2}{3} \right)^6\, N_c^3\, T_0^5~
     \frac{ ( 5 - \kappa ) ( 3 + \kappa ) }
          { \sqrt{1 + \kappa\, }\, ( 3 - \kappa ) }~.
\label{eq:heat_capacity-grand_canonical-M5}
\end{align}
Thus, $\kappa=3$ is the critical line. The static critical exponents $(\alpha,\beta,\gamma,\delta)$ are the same as the $D=5$ case, and $(\nu, \eta)=(3/10,1/3)$ from \eq{static_scaling_exp-III_generic} \cite{Cai:1998ji,Cvetic:1999rb}. 

The equations of motion for $(T,Z,A)$ in the $q = 0$ limit are given by
\begin{subequations}
\label{eq:7dEOMs}
\begin{align}
  & Z'' + \frac{u\, f' - 2f}{u\, f}\, Z'
  + \frac{9\mfw^2\, \bmH}{4\, uf^2}\, Z
  = 0~,
\label{eq:7dEq_Z} \\
  & T' + \frac{\kappa\, u^2}{\bmH}\, A = 0~,
\label{eq:7dconst} \\
  & A'' + \left( \frac{\bmH'}{\bmH} + \frac{f'}{f} - \frac{1}{u}
          \right)^{\mathstrut}\, A'
  + \frac{9\mfw^2\, \bmH}{4u\, f^2}\, A
  + \frac{4u(1 + \kappa)}{f\, \bmH}\, T' = 0~.
\label{eq:7dEq_A}
\end{align}
\end{subequations}
Equations~(\ref{eq:7dEOMs}) give an equation for $A$ only:
\begin{align}
  & A'' + \left( \frac{\bmH'}{\bmH} + \frac{f'}{f} - \frac{1}{u}
          \right)^{\mathstrut}\, A'
  + \left( \frac{9\mfw^2\, \bmH}{4u\,f^2}
  - \frac{4\kappa\, (1 + \kappa)}{f\, \bmH^2}\,u^3 \right)\, A
  = 0~.
\end{align}
Then, the solution is given by
\begin{align}
   A
   &=: a_x^{(0)}\, \frac{f^{-i \mfw T_0/2 T}}{H}\, \calA + O(\omega^2)~, \\
   \calA
  &= 1 + \frac{\kappa u^2}{3}
\nonumber \\
  &+ \frac{i\, \mfw}
          {12 \sqrt{1 + \kappa} (3 + \kappa) (3 + 4 \kappa)^{3/2}}
  \Bigg( 6 (1 + \kappa) (3 + \kappa\, u^2) (2 (\kappa -3) \kappa -9)
  \arctan \frac{\sqrt{3 + 4 \kappa}~u}{2 + u}
\nonumber \\
  &+ \sqrt{3 + 4 \kappa} \Bigg\{ 2 \kappa\, u (3 + \kappa)
  (\kappa (2 \kappa\, u -9)-9) + 9 (1 + \kappa) (3 + 4 \kappa)
  \left( 3 + \kappa u^2 \right) \ln\frac{f}{1-u} \Bigg\} \Bigg)~.
\label{eq:sol_calA-M5}
\end{align}

The boundary action is given by
\begin{align}
   S_{\text{boundary}}
  &= \left(\frac{2}{3}\right)^3 \frac{\pi N_c^{3}\, T_0^4}{9}~\lim_{u \to 0} \int dt\, d\vecx~
    \left( \frac{f\, H}{u} a_x'\, a_x + \cdots \right)~.
\end{align}
Substituting \eq{sol_calA-M5} into the boundary action, we get the correlator, the conductivity, and the diffusion constant:
\begin{align}
  & \tilG^{(R)}_{xx}(\omega, q = 0) = \frac{2^6 \pi\, \kappa\, N_c^3\, T_0^4}{3^6} 
  - i\, \omega \, \frac{4 (3 + \kappa)^2\, N_c^3\, T_0^3}
                       {3^6\, \sqrt{1 + \kappa}}~, \\
  & \lambda = \frac{4 (3 + \kappa)^2\, N_c^3\, T_0^3}
                        {3^6\, \sqrt{1 + \kappa}}~, \\
  & D_R
  = \frac{3}{8 \pi T}~\frac{ (1 + \kappa/3)^3 (1 - \kappa/3) }{(1 + \kappa)(1 + 4\kappa - \kappa^2)}~.
\end{align}
Thus, $\lambda$ is finite and $D_R=0$ on the critical line $\kappa=3$.

\section{Discussion}\label{sec:discussion}

\subsection{A remark on model H}\label{sec:modelH}

We have shown that single R-charge black holes belong to model B.
Model B corresponds to a system with a conserved charge. But if there is another conserved quantity in the system, and if they are coupled to each other, the dynamic universality class becomes model H. 
For our systems, there is an obvious candidate for such a conserved quantity: $T_{\mu\nu}$. In this case, $\lambda$ [defined in \eq{Kubo_formula}] and $\bareta$ become singular as well, and the model is characterized by two dynamic exponents $x_\lambda$ and $x_\eta$:
\bea
D_R &=& \lambda \chi_T^{-1} \sim \xi^{x_\lambda} \chi_T^{-1}~, 
\label{eq:x_lambda-def} \\
\bareta &=& \xi^{x_\eta}~.
\label{eq:x_eta-def}
\eea
These dynamic critical exponents satisfy the following constraint:%
\footnote{Equation~(\ref{eq:H_constraint}) corrects the sign in front of $\eta$ in Ref.~\cite{hohenberg_halperin}.}
\be
x_\lambda + x_\eta = 4 - d - \eta~.
\label{eq:H_constraint}
\ee

However, we found that $\lambda$ remains finite for single R-charge black holes, which suggests model~B. Also, it has been shown that $\bareta$ remains finite for the $D=5$ black hole \cite{Son:2006em,Mas:2006dy}. 

This is  because we focus on linear perturbations. The difference between model B and H comes from a ``mode-mode coupling," which arises from nonlinear terms in hydrodynamic equations. As far as linear perturbations are concerned, there is no distinction between model B and H.

\subsection{The value of the dynamic exponent $z$}\label{sec:dynamic_exponent}

Model B satisfies the dynamic scaling relation $z=4-\eta$ from the theory of dynamic critical phenomena. What is the value of the exponent for single R-charge black holes?
Let us focus on the $D=5$ case.
The value of $z$ requires the knowledge of $\eta$. As discussed in \appen{staticR}, $\eta=1$ from static scaling laws, which implies $z=3$.

There is a caveat though. We extracted the exponents $(\nu, \eta)$ from a static hyperscaling relation, but the hyperscaling relation is often trustable {\it below} the upper critical dimension.
Substituting the values of static exponents \eq{5d_static_exponents} into the Ginzburg criterion (\ref{eq:critical_dim}), one gets $d_c=3$. So, our system is marginal, and the hyperscaling relation may not be trustable. One should not take the value $z=3$ too literally. 

In order to avoid this problem, it is desirable to obtain the relaxation time in the critical regime, which is expected to behave as $1/\tau_q \propto q^z$. In the gravity side, this amounts to solve the scalar mode of the bulk U(1) field, and the relaxation time is given by the lowest quasinormal frequency $\omega_{\text{QNM}}(q)$ of the scalar mode:
\begin{align}
  & 1/\tau_q \sim - \Im\left[~\omega_{\text{QNM}}(q)~\right]~.
\label{eq:tau-QNM}
\end{align}
Thus, the dispersion relation of the quasinormal frequency should behave as
\begin{align}
  & \Im\left[~\omega_{\text{QNM}}(q)~\right]
  \propto q^z~.
\label{eq:QNM-k_z}
\end{align}
It would be interesting to check if the scalar mode computation really gives $z=3$ implied from the static scaling relations \cite{ref:MNO_critical}. Note that only even powers of $q$ appear in the equations of motion, so one expects that only even powers of $q$ also appear in the dispersion relation as well  unless analyticity is broken.

\subsection{Relation to QCD}

Our systems are single R-charge black holes, which are not realistic from the point of view of QCD. Moreover, according to lattice results, the finite temperature transition from hadronic phase to QGP (at zero baryon chemical potential) is not a phase transition but rather a smooth crossover. Thus, one may wonder if the type of phenomena considered in this paper has any relevance to the real QCD.

However, the finite density transition at $T=0$ is believed to be a first-order phase transition, so there must be an end point of the first-order phase transition line somewhere in the phase diagram, which is a critical point. The precise location of the critical point is still unknown, and one of major goals of future RHIC experiments is this critical end point search (See, {\it e.g.}, Ref.~\cite{Stephans:2006tg}). Physics near the critical point has been studied in Refs.~\cite{Stephanov:1998dy,Stephanov:1999zu,Nonaka:2004pg,Asakawa:2008ti} based on the singular behavior of thermodynamic functions. The dynamic critical phenomena have been studied as well \cite{Berdnikov:1999ph,Fujii:2003bz,Son:2004iv}; it was argued that QCD belongs to model H with the dynamic critical exponent for the baryon diffusion constant $z \sim 3$.

\subsection{Future directions}

We have analyzed the dynamic critical phenomena for single R-charge black holes in the linear regime. One interesting direction is to analyze the system {\it in the nonlinear regime.} In this case, the theory of dynamic critical phenomena predicts (genuine) model H. It would be interesting to see if the black hole computation is consistent with model H predictions. A particularly interesting quantity is the exponent $x_\eta$ in \eq{x_eta-def}; if $x_\eta > 0$, the shear viscosity diverges as well. On the other hand, according to the AdS/CFT duality, $\bareta/s = 1/(4\pi)$ in the strong coupling limit for all known examples. Thus, $x_\eta > 0$ means the violation of the universality for $\bareta/s$. (However, it still satisfies the conjectured universal bound for $\bareta/s$ \cite{Kovtun:2004de}, and it is known that $\bareta$ has only a weak critical singularity.)

In an unstable system, the ordering process occurs and the system tends to change to an ordered phase (spinodal decomposition). Because of the second-order phase transition, a single R-charge black hole solution should be replaced by a new bulk solution which corresponds to a new phase. It is important to find such a bulk solution and to analyze the dynamics after the transition. A particularly important question is the dynamics of the interface between two domains. When a system with second-order phase transition is quenched, phase separation occurs and domains are formed with the interface. The interface is described by a kink solution which connects two phases. It would be interesting to find and to analyze the bulk solution which corresponds to the kink. 

In this paper, we have considered only single R-charge black holes, but it would be interesting to extend the analysis to the other black holes. Because there is ample evidence that $\bareta/s = 1/(4\pi)$ in the linear regime, it would be better to consider the other transport coefficients such as the charge diffusion constant, heat conductivity, and the bulk viscosity. Also, we consider only model B and H, but it would be interesting to identify black holes with the other dynamic universality classes.

The analysis such as the one given in this paper may be useful to condensed-matter applications of the AdS/CFT duality. Namely, given a condensed-matter system, the static and dynamic universality classes may be useful to identify the dual black holes. 
For example, it would be interesting to find a bulk black hole solution in the same universality class as the Ginzburg-Landau theory [{\it i.e.}, the same set of critical exponents as Eq.~(\ref{eq:static_exponent_LG})].

\begin{acknowledgments}
We would like to thank Hirotsugu Fujii, Kenji Fukushima, Teiji Kunihiro, Chiho Nonaka, and Hirofumi Wada  for useful discussions. This research was supported in part by the Grant-in-Aid for Scientific Research (20540285) from the Ministry of Education, Culture, Sports, Science and Technology, Japan.
\end{acknowledgments}

%


\appendix

\section{Static critical exponents for the $D=5$ single R-charge black hole}\label{sec:staticR}

The static critical exponents for the single R-charge black holes were obtained in Refs.~\cite{Cai:1998ji,Cvetic:1999rb}, but we include the $D=5$ computation here for completeness.

First, let us approach to the critical line $(T_c, \mu_c)$ with $\mu=\mu_c$. 
Denote the deviation from the critical line by $\eT$. Since $\pi\, T_0 = \mu_c \sqrt{(1+\kappa)/2 \kappa}$, 
\begin{align}
  & \eT := \frac{T}{T_c} - 1
  = \frac{ (2 - \kappa)^2 }{2\, \sqrt{2 \kappa}\,
                     \left( \sqrt{\kappa} + \sqrt{2}\, \right)^2 }
  \propto (2 - \kappa)^2~.
\label{eq:epsT}
\end{align}
Then, thermodynamic quantities behave as
\begin{align}
   C_\mu
  &= \frac{N_c^2\, \mu_c^3}{4\, \pi}\, 
     \frac{ ( 1 + \kappa ) ( 2 + \kappa ) ( 3 - \kappa ) }
          { \sqrt{2\, \kappa}~( 2 - \kappa ) }
  \propto \frac{1}{2 - \kappa} \propto \eT^{-1/2}~,
\label{eq:deviation_T-heat_capacity-grand_canonical-1charge} \\
   \rho - \rho_c
  &= \frac{N_c^2\, \mu_c^3}{32\, \pi^2\, \kappa}\,
  ( \kappa - 2 )\, ( 2\, \kappa - 1 )
  \propto - ( 2 - \kappa ) \propto - \eT^{1/2}~,
\label{eq:deviation_T-rho-1charge} \\
   \chi_T
  &= \frac{N_c^2\, \mu_c^2}{16\, \pi^2}~
     \frac{(1+\kappa)\, (2+5 \kappa-\kappa^2)}{\kappa\, (2-\kappa)}
  \propto \frac{1}{2 - \kappa} \propto \eT^{-1/2}~,
\label{eq:deviation_T-electric_susceptibility-1charge}
\end{align}
which determine 3 static critical exponents
\begin{align}
  & \alpha = \beta = \gamma = 1/2~.
\label{eq:static_scaling_exp-I}
\end{align}

Similarly, let us approach the critical line with $T=T_c$, and
denote the deviation from the critical line by $\epsmu$.
Since $T_0 = 2\, T_c\, \sqrt{1+\kappa}/(2+\kappa)$,
\begin{align}
  & \epsmu := 1 - \frac{\mu}{\mu_c}
  = \frac{ (2 - \kappa)^2 }{ (2 + \kappa)\,
                     \left( \sqrt{\kappa} + \sqrt{2}\, \right)^2 }
  \propto (2 - \kappa)^2~.
\label{eq:epsmu}
\end{align}
The R-charge density then behaves as
\begin{align}
   \rho - \rho_c
  &= \frac{\pi\, N_c^2\, T_c^3}{4}
  \frac{ 4\, (1+\kappa)^2\, \sqrt{2\, \kappa} - 9\, (2+\kappa)^3 }
       { (2+\kappa)^3 }
  \propto - ( 2 - \kappa ) \propto - \epsmu^{1/2}~,
\label{eq:deviation_mu-rho-1charge}
\end{align}
which determines $\delta$ as
\begin{align}
  & \delta = 2~.
\label{eq:static_scaling_exp-II}
\end{align}

The static critical exponents (\ref{eq:static_scaling_exp-I}) and 
(\ref{eq:static_scaling_exp-II}) satisfy the static scaling relations
\begin{align}
  & \alpha + 2\, \beta + \gamma = 2~,
& & \gamma = \beta \left( \delta - 1 \right)~.
\label{eq:static_scaling_law-thermo}
\end{align}
The static exponents $\nu$ and $\eta$ require the R-charge correlator, but it has not been evaluated.  Instead, if we use the scaling relations
\begin{align}
  & 2 - \alpha = \nu~d~,
& & \gamma = \nu \left( 2 - \eta \right)~,
\label{eq:static_scaling_law-Green}
\end{align}
we get
\be
(\nu, \eta) = \left( \frac{3}{2d}\, , \frac{6-d}{3} \right)~,
\label{eq:static_scaling_exp-III_generic}
\ee
or for $d=3$, 
\be
(\nu, \eta) = \left( \frac{1}{2}\, , 1 \right)~.
\label{eq:static_scaling_exp-III}
\ee
Note that the value of $\eta$ differs from the usual mean-field theory value $\eta = 0$.

Consequently, we get
\be
(\alpha, \beta, \gamma, \delta, \nu, \eta) = \left(\frac{1}{2}, \frac{1}{2}, \frac{1}{2}, 2,  \frac{1}{2}, 1\right)~.
\ee


\end{document}